\documentclass[11pt]{article} 
\usepackage{graphicx,amssymb,mathrsfs,array,multirow}  
\begin{document}  

\vskip 30pt

\begin{center}  
{\Large{\bf Are the small neutrino oscillation parameters all related? }}\\
\vspace*{1cm}  
\renewcommand{\thefootnote}{\fnsymbol{footnote}}  
{ {\sf Soumita Pramanick\footnote{soumitapramanick5@gmail.com},} 
{\sf Amitava Raychaudhuri\footnote{palitprof@gmail.com}}
} \\  
\vspace{10pt}  
{\small  {\em Department of Physics, University of Calcutta,  
92 Acharya Prafulla Chandra Road, Kolkata 700009, India}}
\normalsize

\end{center}  

\begin{abstract} 

Neutrino oscillations reveal several small parameters, namely,
$\theta_{13}$, the solar mass splitting {\em vis-\`{a}-vis} the
atmospheric one, and the deviation of $\theta_{23}$ from maximal
mixing. Can these small quantities all be traced to a single
source and, if so, how could that be tested? Here a  see-saw model for
neutrino masses is presented wherein  a dominant term
generates the atmospheric mass splitting with maximal mixing in
this sector, keeping $\theta_{13} = 0$ and zero solar splitting.
A Type-I see-saw perturbative contribution results in  non-zero
values of $\theta_{13}$, $\Delta m^2_{solar}$, $\theta_{12}$,  as
well as allows $\theta_{23}$ to deviate from $\pi/4$ in
consistency with the data while interrelating them all.
CP-violation is a natural consequence and is large ($\delta \sim
\pi/2, 3\pi/2$) for inverted
mass ordering.   The model will be tested as
precision on the neutrino parameters is sharpened.

\vskip 5pt 
\noindent  
\texttt{Key Words:~~Neutrino mixing, $\theta_{13}$, Leptonic CP-violation,
Neutrino Mass ordering, Perturbation}
\end{abstract}  

\renewcommand{\thesection}{\Roman{section}} 
\setcounter{footnote}{0} 
\renewcommand{\thefootnote}{\arabic{footnote}} 
\noindent


Information on neutrino mass and mixing have been steadily
emerging from oscillation experiments.  Among them the
angle\footnote{For the lepton mixing matrix the standard PMNS
form is used.}
$\theta_{13}$ is small ($\sin \theta_{13} \sim 0.1$) \cite{t13} while global
fits to the solar, atmospheric, accelerator, and reactor neutrino
oscillation data indicate that $\theta_{23}$ is near maximal ($\sim
\pi/4$) \cite{Gonzalez, Valle}. On the other hand, the solar
mass square difference is two orders smaller than the atmospheric
one.  These mixing parameters and the mass ordering are essential
inputs for identifying viable models for neutrino masses. 

A natural choice could be to take the mixing angles to be
initially either
$\pi/4$ ($\theta_{23}$) or zero ($\theta_{13}$, $\theta_{12}$)
and the solar splitting absent. In this spirit, here  a
proposal is put forward under which the atmospheric mass
splitting and maximal mixing in this sector arise from a
zero-order mass matrix while   the smaller solar mass splitting
and  realistic  $\theta_{13}$ and $\theta_{23}$  are generated by
a Type-I see-saw \cite{seesaw} which acts as a perturbation.
$\theta_{12}$ also arises out of the same perturbation and as a
consequence of degeneracy is not constrained to be small.
Attempts to generate {\em some} of the neutrino parameters by
perturbation theory are not new
\cite{old, pert}, but to our knowledge there is no work in the
literature that indicates that {\em all} the small parameters
could have the same perturbative origin  and agree with the
current data.

The unperturbed neutrino mass matrix in the mass basis is 
$M^0 = {\rm diag} \{ m^{(0)}_1, m^{(0)}_1, m^{(0)}_3 \}$
with the mixing matrix of the form
\begin{equation}
U^0= \pmatrix{1 & 0 & 0 \cr
0 & \sqrt{1 \over 2} & \sqrt{1 \over 2} \cr
0 & -\sqrt{1 \over 2} & \sqrt{1 \over 2}} \;\;.
\label{mix0}
\end{equation}
Here $\Delta m^2_{atm} = (m^{(0)}_3)^2 - (m^{(0)}_1)^2$. By
suitably choosing the Majorana phases the
masses $m^{(0)}_1, m^{(0)}_3$ are taken to be real and positive.
The columns of $U^0$ are the unperturbed flavour
eigenstates\footnote{In the flavour basis the charged lepton mass
matrix is diagonal.}. As stated, $\Delta m^2_{solar} = 0$ and
$\theta_{13} = 0$. Since the first two states are degenerate in
mass, one can also take $\theta_{12} =0$.
It is possible to generate this mass matrix from a Type-II
see-saw.

In the flavour basis the mass matrix is $(M^{0})^{flavour} = U^{0}
M^{0} U^{0T}$ which in terms of  
$m^{\pm} = m_3^{(0)} \pm m_1^{(0)}$ is
\begin{equation}
(M^{0})^{flavour} = {1 \over 2}
\pmatrix{2m^{(0)}_1 & 0 & 0 \cr 0 & m^+ & m^- \cr 0 & m^- & m^+} 
\;\;.
\label{mflav0}
\end{equation}

The perturbation is obtained by a Type-I see-saw.  To reduce the
number of independent parameters, in the
flavour
basis the Dirac mass term is taken to be proportional to the
identity, i.e., 
\begin{equation}
M_D = m_D ~\mathbb{I} \;.
\end{equation}
In this basis, in the
interest of minimality the
right-handed neutrino Majorana mass matrix is taken with only two
non-zero complex entries.
\begin{equation}
 M_R^{flavour} ={m_R} 
\pmatrix{0 & x e^{-i\phi_1} & 0 \cr x e^{-i\phi_1} & 0 & 0 \cr 0 & 0
& y e^{-i\phi_2}}  \  \ ,
\label{mflav1}
\end{equation}
where $x, y$ are dimensionless constants of ${\cal O}(1)$. No
generality is lost by keeping the Dirac mass real.

As a warm-up consider first the real case, i.e., $\phi_1 = 0
~{\rm or} ~\pi, \phi_2 = 0 ~{\rm or} ~\pi$. For notational
convenience in the following the phase factors are not
displayed; instead $x$ ($y$) is taken as positive or
negative depending on whether $\phi_1$ ($\phi_2$) is 0 or $\pi$.
Negative $x$ and $y$ offer interesting variants which are
stressed at the appropriate  points.

The Type-I see-saw contribution in the mass basis is:
\begin{equation}
 M'^{mass}
= U^{0T} \left[M_D^T(M_R^{flavour})^{-1}M_D \right] U^0 = 
{m_D^2 \over \sqrt 2 ~xy m_R} 
\pmatrix{0 & y & y \cr y & {x \over  \sqrt 2} & -{x \over \sqrt 2} \cr 
y & -{x \over \sqrt 2} & {x \over \sqrt 2}}\;\;.
\label{pert1}
\end{equation}

The effect on the solar sector is governed by the submatrix of
$M'^{mass}$ in the subspace of the two degenerate states, 
\begin{equation}
M'^{mass}_{2\times2} = {m_D^2 \over \sqrt 2 ~xy m_R} 
\pmatrix{0 & y \cr y & {x/\sqrt 2}} \;.
\label{solr}
\end{equation}
To first order in the
perturbation:
\begin{equation}
\tan 2\theta_{12}= 2 \sqrt 2 \left(\frac{y}{x}\right)  \; .
\label{solangr}
\end{equation}
For $y/x = 1$ one obtains the tribimaximal mixing  value of
$\theta_{12}$ which, though allowed by the data\footnote{We use
the  3$\sigma$ ranges $7.03  \leq \Delta m_{21}^2/ 10^{-5} \, {\rm
eV}^2 \leq 8.03 \;\; {\rm and}\;\; 31.30^\circ \leq \theta_{12}
\leq 35.90^\circ$ \cite{Gonzalez}.} at 3$\sigma$, is beyond the
$1\sigma$ region.  Since for the entire range of $\theta_{12}$
one has  $\tan 2\theta_{12} > 0$, $x$ and $y$ must be chosen of
the same sign.   Therefore, either $\phi_1 = 0 = \phi_2$ or
$\phi_1 = \pi = \phi_2$. From the global fits to the experimental
results one finds:
\begin{equation}
0.682 < \frac{y}{x} < 1.075 ~{\rm at} ~3\sigma \;.
\label{t12lim}
\end{equation}
Further, from eq. (\ref{solr}),
\begin{equation}
\Delta m^2_{solar}=  {m_D^2 \over xy m_R} ~m^{(0)}_1
\sqrt{x^2 + 8y^2}\;. 
\label{solspltr}
\end{equation} 

To first order in the perturbation the corrected
wave function $|\psi_3\rangle$ is:
\begin{equation}
|\psi_3\rangle =
\pmatrix{\kappa \cr {1\over \sqrt 2}{(1 - \frac{\kappa}{\sqrt 2} {x\over y})} 
\cr {1\over \sqrt 2}{(1 + \frac{\kappa}{\sqrt 2} {x\over y})} } \ \ \ ,
\label{psi3_1}
\end{equation} 
where
\begin{equation}
\kappa \equiv {m_D^2 \over {\sqrt 2 ~x m_R m^-}} \;\;.
\label{kappa}
\end{equation} 
For positive $x$  the sign of $\kappa$ is fixed by that of $m^-$.
Since by convention  all the mixing angles $\theta_{ij}$ are in
the first quadrant, from eq.  (\ref{psi3_1}) one must identify:
\begin{equation}
\sin \theta_{13}\cos\delta=\kappa={m_D^2 \over {\sqrt 2 ~x m_R m^-}}\;\;,
\label{s13}
\end{equation}
where for $x > 0$  the PMNS phase $\delta = 0$ for normal mass ordering
(NO) and $\delta = \pi$ for inverted mass ordering
(IO).
Needless to say, both these cases are CP conserving. 
If $x$ is negative then NO (IO) would
correspond to $\delta = \pi ~(0)$.
 
An immediate consequence of eqs. (\ref{s13}), (\ref{solangr}), and
(\ref{solspltr}) is
\begin{equation}
\Delta m^2_{solar} =  {\rm sgn}(x)~m^- m^{(0)}_1 
~\frac{4 \sin \theta_{13} \cos\delta}{\sin 2\theta_{12}}  \;, 
\label{solsplr2}
\end{equation}
which exhibits how the solar sector and $\theta_{13}$ are intertwined. 
The positive  sign of $\Delta m^2_{solar}$, preferred by the data,
is  trivially verified since  ${\rm sgn}(x)~m^- \sin
\theta_{13} \cos\delta > 0$  from eq. (\ref{s13}).
However, eq. (\ref{solsplr2})   excludes  inverted
ordering.  Once the neutrino mass square splittings,
$\theta_{12}$,  
and $\theta_{13}$ are chosen, eq.
(\ref{solsplr2}) determines the lightest neutrino mass, $m_0$.
Defining $z = m^- m^{(0)}_1/\Delta m^2_{atm}$ and
$ m_0/\sqrt{|\Delta m^2_{atm}|} =  \tan \xi$, one has
\begin{eqnarray}
z &=& \sin \xi/(1+ \sin \xi) \;\; 
{\rm (normal ~ordering)},\nonumber \\
z &=& 1/(1+ \sin \xi) \;\; {\rm (inverted  ~ordering)} \;\;. 
\label{m_0}
\end{eqnarray} 
It is seen  that $0 \leq z \leq 1/2$ for NO
and $1/2 \leq z \leq 1$ for IO,
with $z \rightarrow 1/2$ corresponding to quasidegeneracy,
i.e., $m_0 \rightarrow $ large, in both
cases.  From eq. (\ref{solsplr2}) 
\begin{equation}
z = \left(\frac{\Delta m^2_{solar}}{|\Delta m^2_{atm}|}\right) 
\left(\frac{\sin 2\theta_{12}}{4 \sin \theta_{13} |\cos\delta|}\right) \;\;,
\label{z}
\end{equation} 
with $|\cos\delta| = 1$ for real $M_R$. As shown below,  the
allowed ranges of the oscillation parameters imply $z
\sim 10^{-2}$ and so inverted mass
ordering is disallowed.

From eq. (\ref{psi3_1}) one further finds: 
\begin{equation}
\tan\theta_{23} \equiv \tan (\pi/4 - \omega) = \frac{1-\frac{\kappa}{\sqrt 2} {x\over y}}  
{1 + \frac{\kappa}{\sqrt 2} {x\over y}}   
\ \ , 
\label{th23r}
\end{equation}
where, using eqs. (\ref{solangr}) and (\ref{s13}), 
\begin{equation}
\tan \omega = 
\frac{2 \sin \theta_{13}\cos\delta}{\tan2\theta_{12}} \;.
\label{phir}
\end{equation}
$\theta_{23}$ will be in the first (second) octant, i.e., 
the sign of $\omega$ will be
positive (negative) if $\delta = 0 ~(\pi)$.  
Recall, this corresponds to $x > 0$ ($x < 0$).

\begin{figure}[tb] 
\begin{center} 
{\includegraphics[scale=0.6,angle=0]{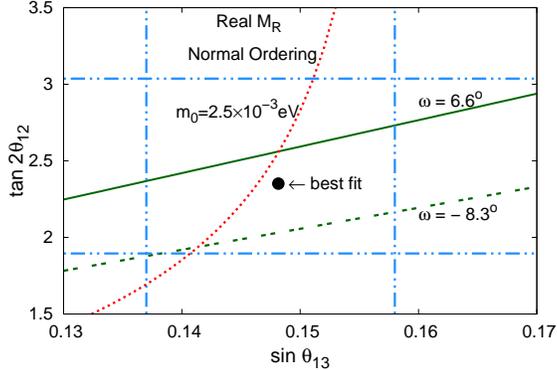}}
\caption{ \sf  \small The blue dot-dashed box is
the global-fit 3$\sigma$ allowed range of $\sin \theta_{13}$ and $\tan
2\theta_{12}$. The best-fit point is shown as a black dot. The red
dotted curve is from eq.
(\ref{solsplr2}) with $m_0 = 2.5$ meV when the best-fit values of the two
mass-splittings are used. 
The portion below the green solid (dashed) straight line  is excluded by
$\theta_{23}$ at 3$\sigma$ -- eq. (\ref{phir}) -- for the first
(second) octant. In case of inverted ordering no solution
of eq. (\ref{solsplr2}) is allowed for real $M_R$. }
\label{Real1} 
\end{center} 
\end{figure} 

In Fig. \ref{Real1} the global-fit 3$\sigma$ range of $\sin
\theta_{13}$ and $\tan 2\theta_{12}$ is shown as the blue
dot-dashed box with the best-fit value indicated by a black dot.
Once the atmospheric and solar mass splittings are fixed, for any
point within this region  eq.  (\ref{z}) determines a  $z$, or
equivalently an $m_0$, which leads to the correct solar
splitting.

From the  3$\sigma$ data \cite{Gonzalez} $\omega_{min} = 0$ for both
octants and $\omega_{max} = 6.6^\circ ~(-8.3^\circ)$ for the
first (second) octant. As $|\cos\delta | = 1$ for
the real $M_R$ case, in this model one has from eq.  (\ref{phir})
for both octants $|\omega| \geq 5.14^\circ$ at 3$\sigma$.  Thus
the range of $\theta_{23}$ that can be obtained  is rather
limited\footnote{This range is excluded at 1$\sigma$ for the first
octant.}. The green solid (dashed)  straight line is from eq.
(\ref{phir}) for $\omega_{max}$ for the first (second) octant.
The region {\em below} this line is  excluded in this model.
Note that the best-fit point  is permitted only for $\theta_{23}$
in the second octant.

Using the 3$\sigma$ global-fit limits of $\theta_{13}$ and
$\theta_{12}$, from eq. (\ref{z}) one gets $z_{max} = 6.03 \times
10^{-2}$ implying that $(m_0)_{max}$ = 3.10 meV.   Also, consistency with
both eqs.  (\ref{phir}) at $\omega_{max}$ and (\ref{z}) sets
$z_{min}$ = 4.01 $\times 10^{-2}$ (3.88 $\times 10^{-2}$) for the
first (second) octant corresponding to $(m_0)_{min}$ = 2.13
(2.06) meV.  If, as a typical example, $m_0 = 2.5$ meV is taken and
the best-fit values of the solar and atmospheric mass splittings
are used then eq. (\ref{solsplr2}) gives the red dotted curve in
Fig. \ref{Real1}.

In summary, for  real $M_R$ the free parameters are $m_0$,
$m_D^2/xm_R$ and $y$ with which the solar mass splitting,
$\theta_{12}, \theta_{13}, \theta_{23}$ are reproduced for normal
mass ordering. Inverted ordering cannot be accommodated.

Reverting now to  the complex  $M_R$ in eq. (\ref{mflav1})
one has in the mass basis in place of eq. (\ref{pert1}):
\begin{equation}
 M'^{mass} =
{m_D^2 \over \sqrt 2 x y m_R} 
\pmatrix{0 & y e^{i\phi_1} & y e^{i\phi_1} \cr y e^{i\phi_1} &
x {e^{i\phi_2}}\over{\sqrt{2}} & - x {e^{i\phi_2}}\over{\sqrt{2}} \cr
y e^{i\phi_1} & -{x e^{i\phi_2}}\over{\sqrt{2}}&
x {e^{i\phi_2}}\over{\sqrt{2}}}.
\label{TypeIcmplxM}
\end{equation}

$x$ and $y$ are now positive. $M'$ is no longer hermitian. This is 
addressed, as usual, by defining the hermitian combination $(M^0
+ M')^\dagger(M^0 + M')$ and treating $M^{0\dagger} M^0$ as the
unperturbed term and $(M^{0\dagger} M' + M'^\dagger M^0)$ as the
perturbation to lowest order. The zero order eigenvalues are now
$(m^{(0)}_i)^2$ and the complex yet hermitian perturbation matrix
is
\begin{equation}
(M^{0\dagger} M' + M'^\dagger M^0)^{mass} = 
{m_D^2 \over \sqrt 2 xy m_R}\pmatrix{ 0 & 2 m^{(0)}_1 y \cos\phi_1 & 
y f(\phi_1) \cr
2 m^{(0)}_1 y \cos\phi_1 & { 2 \over \sqrt{2}}m^{(0)}_1 x \cos\phi_2 & 
-{1\over\sqrt{2}} x f(\phi_2)\cr
y f^*(\phi_1)& 
 -{1\over\sqrt{2}} x f^*(\phi_2)& 
{ 2 \over \sqrt{2}}m^{(0)}_3 x \cos\phi_2},
\label{pertcmplx}
\end{equation}
where
\begin{equation}
f(\xi) = m^{+} \cos\xi - i m^{-} \sin\xi \;\;.
\label{ffn}
\end{equation}
The subsequent analysis is similar to the one for real $M_R$.

The perturbation which splits the degenerate solar sector is the
$2\times2$ block of eq. (\ref{pertcmplx}).
The solar mixing angle now is 
\begin{equation}
\tan 2\theta_{12}= 2\sqrt2 ~{y \over x}
~{\cos\phi_1\over\cos\phi_2} \;.
\label{solangcmplx}
\end{equation}
The limits
of eq. (\ref{t12lim}) apply on the ratio $(y \cos\phi_1/x \cos\phi_2)$. Also,
$(\cos\phi_1/\cos\phi_2)$ must be positive. 

Including first order corrections the wave function
$|\psi_3\rangle$  is 
\begin{equation}
|\psi_3\rangle =
\pmatrix{\kappa f(\phi_1)/m^+\cr {1\over \sqrt 2}{(1-{\kappa \over \sqrt
2}\frac{x}{y} ~f(\phi_2)/m^+)} \cr {1\over \sqrt 2}{(1+{\kappa\over \sqrt
2}  \frac{x}{y} ~f(\phi_2)/m^+)}
} .
\label{psi3ca}
\end{equation}
Now $\kappa$ is positive (negative) for NO (IO). One immediately has
\begin{eqnarray}
\sin \theta_{13}\cos\delta &=& \kappa  \cos\phi_1 \ , 
\nonumber \\ 
\sin \theta_{13}\sin\delta &=& \kappa ~\frac{m^-}{m^+} \sin\phi_1\  \ . 
\label{s13cmplx}
\end{eqnarray}
The sign of $\cos\delta$ is the same as (opposite of)  sgn$(\cos
\phi_1)$ for normal (inverted) mass ordering. Further, $\sin
\phi_1$ determines the combination $\sin \theta_{13} \sin\delta$
that appears in the Jarlskog parameter, $J$, a measure of
CP-violation.  Note, $\phi_2$ plays no role in fixing the CP-phase
$\delta$.

It is seen that for normal ordering ($\kappa > 0$) the quadrant
of $\delta$ is the same as that of $\phi_1$. For inverted
ordering ($\kappa < 0$)  $\delta$ is in the first (third)
quadrant if $\phi_1$ is in the second (fourth) quadrant and {\em
vice-versa}.

$\theta_{23}$ obtained from eq. (\ref{psi3ca}) is
\begin{equation}
\tan\theta_{23} = {{1-{\kappa \over
\sqrt 2}\frac{x}{y}\cos\phi_2} \over  {1+{\kappa\over \sqrt
2}\frac{x}{y} \cos\phi_2}}  \ \ \ \ ,
\end{equation}
where, using eqs. (\ref{solangcmplx}) and (\ref{s13cmplx}),
\begin{equation}
\tan\omega 
= \frac{2\sin \theta_{13}\cos\delta}{\tan2\theta_{12}} \;.
\label{phic}
\end{equation}
Eq. (\ref{phir}) is recovered when $\cos
\delta = \pm 1$. 
From eq. (\ref{phic}), if $\delta$ lies in the first or the fourth quadrant --
which yield opposite signs of $J$ -- 
$\theta_{23}$ is in the first octant
while it is in the second octant otherwise.

A straight-forward calculation 
after  expressing $m_D$ and $m_R$ in terms of
$\sin \theta_{13}\cos\delta$, yields
\begin{equation}
\Delta m^2_{solar}
=  {\rm sgn}(\cos\phi_2) ~m^- m^{(0)}_1 
~\frac{4 \sin \theta_{13} \cos\delta }{\sin 2\theta_{12}} 
\;,
\label{solsplc}
\end{equation}
which bears a strong similarity with eq. (\ref{solsplr2}) for
real $M_R$. Eqs. (\ref{m_0}) and (\ref{z}) continue to hold. 
To ensure the positivity of $\Delta m^2_{solar}$, noting the
factors determining  the sign of $\cos\delta$, one concludes that
sgn$(\cos\phi_1
\cos\phi_2)$ must be positive for both mass orderings. Thus, satisfying the
solar mass splitting leaves room for either octant of
$\theta_{23}$ for both mass orderings.
The allowed range of $\delta$ can be easily read off if we
reexpress eq. (\ref{z}) as:
\begin{equation}
|\cos\delta| = \left(\frac{\Delta m^2_{solar}}{|\Delta m^2_{atm}|}\right) 
\left(\frac{\sin 2\theta_{12}}{4 \sin \theta_{13} ~z}\right) \;\;.
\label{cdel}
\end{equation}

\begin{figure}[tb] 
\begin{center} 
{\includegraphics[scale=0.6,angle=0]{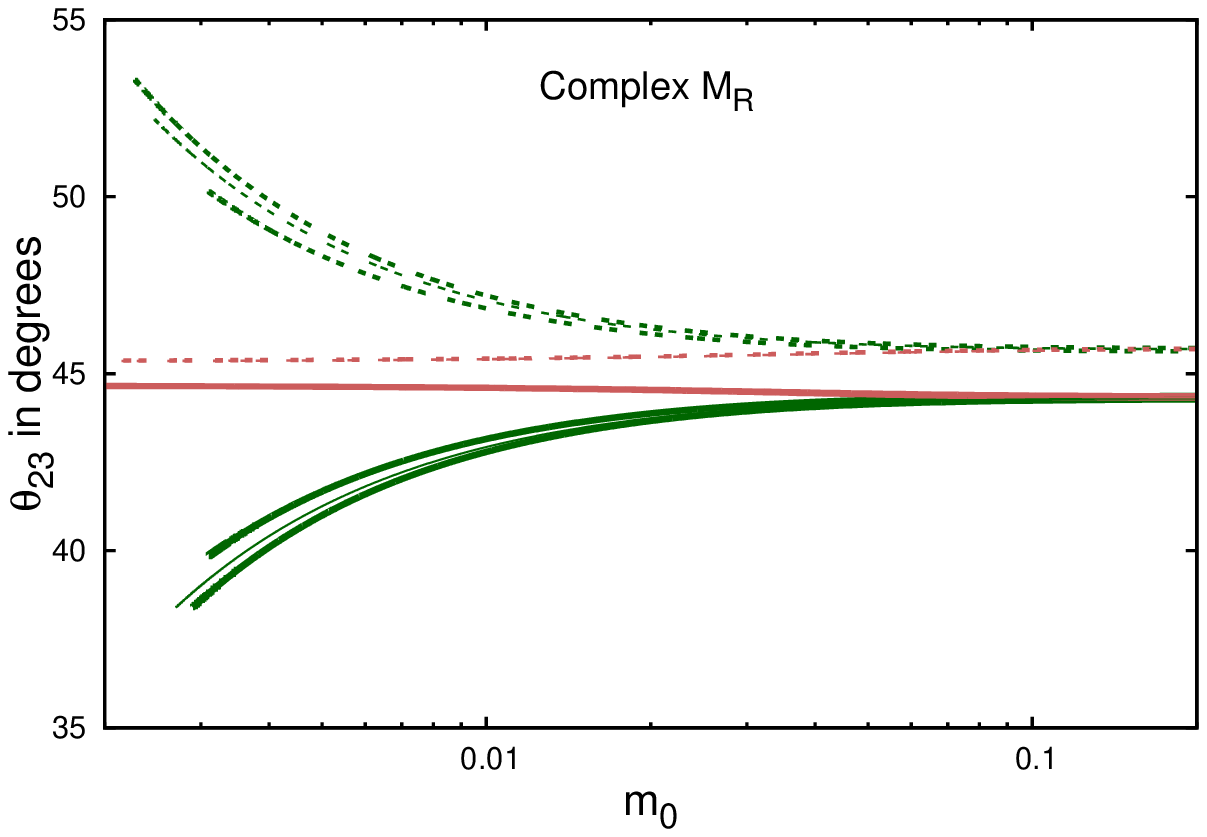}
\includegraphics[scale=0.6,angle=0]{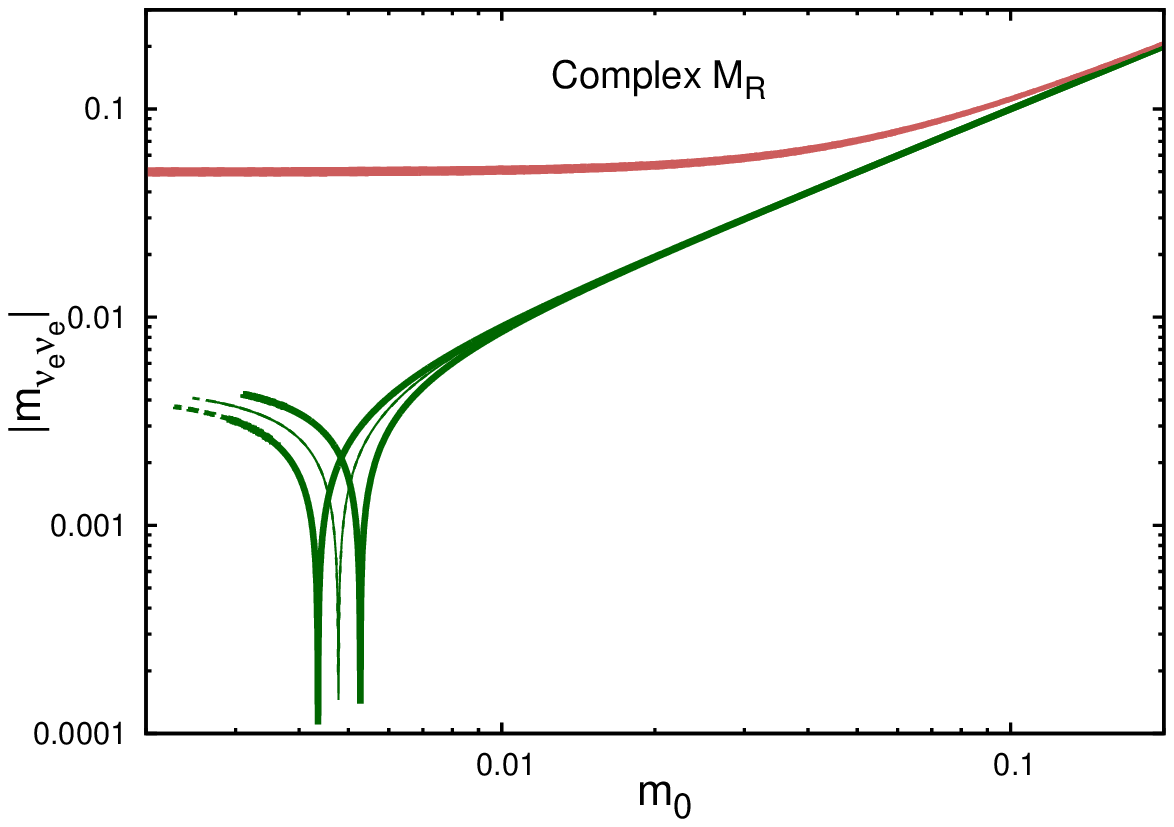}}
\caption{\sf \small $\theta_{23}$ ($|m_{\nu_e \nu_e}|$  in  eV)
as a function of the lightest neutrino mass $m_0$ (in  eV) is
shown in the left (right) panel.  The
green (pink) curves are for the normal (inverted) mass ordering.
For every plot the region allowed at 3$\sigma$ is between the
thick curves while the thin curves are for the best-fit values of
the inputs. The solid (dashed) curves correspond to the first
(second) octant of $\theta_{23}$.    }
\label{th23} 
\end{center} 
\end{figure} 

\begin{figure}[tb] 
\begin{center} 
{\includegraphics[scale=0.6,angle=0]{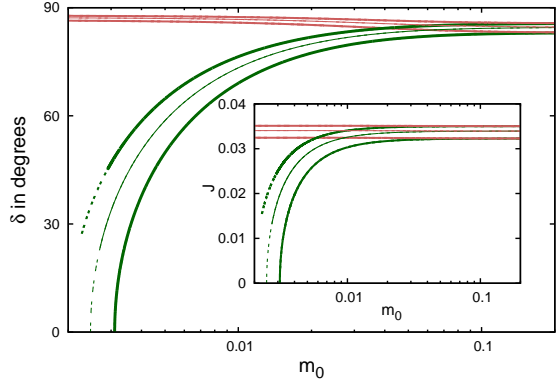}}
\caption{\sf \small The CP-phase $\delta$ is plotted
as a function of $m_0$ (in  eV).  Inset:  The leptonic
CP-violation measure $J$ is shown. The
conventions are the same as that of Fig. \ref{th23}. }
\label{CP} 
\end{center} 
\end{figure} 

In the following $m_0,  \theta_{13}$, and $\theta_{12}$ are taken
as inputs and $\delta$ and $\theta_{23}$ are obtained using
eqs. (\ref{cdel}) and (\ref{phic}). 
From these the CP-violation
measure, $J$, and the combination  $|m_{\nu_e \nu_e}|$
which determines the rate of neutrinoless double beta decay 
are calculated. 

In the left panel of Fig. \ref{th23} is shown (thick curves)
the dependence of $\theta_{23}$ on the lightest neutrino mass
$m_0$  when the neutrino mass square splittings  and the angles
$\theta_{13}$ and $\theta_{12}$ are varied over their allowed
ranges at 3$\sigma$. The thin curves correspond to taking the
best-fit values. The green (pink)
curves are for normal (inverted) mass ordering while solid
(dashed) curves are for solutions in the first (second) octant.
For inverted ordering the thick and thin curves are very close and cannot be
distinguished in this figure. Notice that the 3$\sigma$ predictions from this
model are not consistent with $\theta_{23} = \pi/4$.  As expected
from eq.  (\ref{phic}), $\theta_{23}$ values are symmetrically
distributed around $\pi/4$. Its range for inverted ordering falls
outside the 1$\sigma$ global fits but are consistent at
3$\sigma$. An improvement in the determination of $\theta_{23}$
will be the easiest way to exclude one of the orderings unless
one is in the quasidegenerate regime. 
For normal
ordering the smallest value of $m_0$ is determined by the
3$\sigma$ limits of $\theta_{23}$ in the two octants. Eq.
(\ref{solsplc}) permits arbitrarily small  $m_0$ for inverted
mass ordering (see below).  

In the right panel of Fig. \ref{th23} 
$|m_{\nu_e \nu_e}|$
has been plotted.  
The sensitivity of  direct neutrino mass measurements is
expected to reach around 200 meV \cite{katrin} in the near
future.  Planned neutrinoless double beta decay experiments will
also probe the quasidegenerate range of $m_0$ \cite{0nubeta}.  As
can be seen from this figure, to distinguish the two mass
orderings at least a further one order improvement in sensitivity
will be needed. Long baseline experiments or large atmospheric
neutrino detectors such as INO will settle the mass ordering
more readily.

In  Fig. \ref{CP} is displayed the variation of $\delta$ with
$m_0$  for both mass orderings while $J$ is shown in the inset.
The conventions are the same as in Fig. \ref{th23}.  The
sign of $J$ is positive if $\delta$ is in the first or second
quadrant and is negative for the other cases. 
As noted, the quadrant of
$\delta$ (and the associated sign of $J$) can be altered by the
choice of the quadrant of $\phi_1$. However, from eq.
(\ref{cdel}) for these alternatives, namely, $\pm \delta$ and 
$(\pi \pm \delta)$,
the dependence of $|\cos\delta|$ on $m_0$ is the same
for a particular mass ordering.
With this proviso in mind, 
Fig. \ref{CP}  has been plotted keeping $\delta$ in the first
quadrant and $J$ has been taken as positive. 

$J$, which is proportional to $\sin 2\theta_{23}$, has no
dependence on the  octant of $\theta_{23}$ as the latter is
symmetrical around $\pi/4$.  In both Figs. \ref{th23} and \ref{CP}, for normal
ordering a slightly larger range of $m_0$ is allowed when
$\theta_{23}$ is in the second octant. For the region where both
octants are allowed the curves in Fig. \ref{CP} completely
overlap. For inverted mass ordering both $\delta$ and $J$ remain
nearly independent of $m_0$.

For $m_0$ smaller than 10 meV, the CP-phase
$\delta$ is significantly larger for inverted
ordering\footnote{In fact for inverted ordering $\delta$ remains
close to
$\pi/2$ or $3\pi/2$ for all $m_0$.}. This
could provide a clear test of this model when the mass
ordering is known and CP-violation in the neutrino sector is
measured.  The real limit ($\delta$ = 0) is seen to be
admissible, as expected from Fig. \ref{Real1}, only for normal
ordering and that too not for the entire 3$\sigma$ range, with
the second octant allowing a larger region.

Since $0 \leq z \leq  1/2$ for NO and $1/2 \leq z \leq 1$ for IO,
the allowed   values of $\delta$ in the two orderings as seen
from eq. (\ref{cdel}) are
complementary tending towards a common value as $z \rightarrow
1/2$, the quasidegenerate limit, which begins to set in from
around $m_0$ = 100 meV.  The main novelty from
the real $M_R$ case is that in eq.  (\ref{z}) by choosing
$\cos\delta$ sufficiently small one can make $z \equiv m^-
m^{(0)}_1/\Delta m^2_{atm} \sim 1$ so that solutions exist for
$m_0$ for inverted mass ordering   corresponding to even vanishing
$m_0$ unlike the case of normal ordering where the lower limit of
$m_0$ is set by $\cos\delta = 1$, i.e., real $M_R$. 

We have checked that the size of the perturbation is at most
around 20\%
of the unperturbed contribution for all cases.

In conclusion, a model for neutrino masses has been proposed in
which the atmospheric mass splitting  together with
$\theta_{23} = \pi/4$ has an origin  different from that of the
solar mass splitting, $\theta_{12}$, $\theta_{13}$, and  $\omega
= \pi/4 - \theta_{23}$, all of which arise  from a single
perturbation  resulting from a Type-I see-saw.  The global
fits to the mass splittings, $\theta_{12}$ and $\theta_{13}$
completely pin-down the model and the CP-phase $\delta$ and the
octant of $\theta_{23}$ are predicted  in terms of the
lightest neutrino mass $m_0$. Both mass orderings are allowed,
the inverted ordering being associated with near-maximal
CP-violation. Both octants of $\theta_{23}$ can be accommodated.
Further improvements in the determination of $\theta_{23}$, a
measurement of the CP-phase $\delta$, along with a knowledge of
the neutrino mass ordering will put this model to tests from
several directions.

{\bf Acknowledgements:} SP acknowledges a Senior Research
Fellowship from CSIR, India.  AR is partially funded by  the
Department of Science and Technology Grant No. SR/S2/JCB-14/2009.



\begin{thebibliography}{100} 



\bibitem{t13}
For the present status of $\theta_{13}$ see presentations from
Double Chooz, RENO, Daya Bay, MINOS/MINOS+ and T2K at Neutrino
2014.
https://indico.fnal.gov/conferenceOtherViews.py?view=standard\&confId=8022.



\bibitem{Gonzalez} 
  M.~C.~Gonzalez-Garcia, M.~Maltoni, J.~Salvado and T.~Schwetz,
  JHEP {\bf 1212}, 123 (2012)
  [arXiv:1209.3023v3 [hep-ph]], NuFIT 1.3 (2014).

\bibitem{Valle}
  D.~V.~Forero, M.~Tortola and J.~W.~F.~Valle,
  Phys.\ Rev.\ D {\bf 86}, 073012 (2012)
  [arXiv:1205.4018 [hep-ph]].


\bibitem{seesaw}
P. Minkowski,  Phys.\ Lett.\ B {\bf 67}, 421 (1977); 
M.~Gell-Mann, P.~Ramond and R.~Slansky,
in \textit{Supergravity}, p.~315, edited by F. van Nieuwenhuizen 
and D. Freedman, North Holland, Amsterdam, (1979); 
T.~Yanagida, Proc. of the \textit{Workshop on Unified Theory and 
the Baryon Number of the  Universe}, KEK, Japan, (1979); 
S.L. Glashow,  NATO Sci.\ Ser.\ B {\bf 59}, 687 (1980); 
R.N. Mohapatra and G.~Senjanovi{\'c},  Phys.\ Rev.\ D {\bf 23}, 165 (1981).

\bibitem{old} Earlier work on neutrino mass models in which a few
elements dominate over others  can be traced to 
  F.~Vissani,
JHEP {\bf 9811}, 025 (1998)
[hep-ph/9810435].
Models with  somewhat similar points of view as those
espoused here are 
  E.~K.~Akhmedov,
Phys.\ Lett.\ B {\bf 467}, 95 (1999)
[hep-ph/9909217], and 
  M.~Lindner and W.~Rodejohann,
JHEP {\bf 0705}, 089 (2007)
[hep-ph/0703171].

\bibitem{pert} For more recent work after the determination of
$\theta_{13}$ see, for example, 
  B.~Brahmachari and A.~Raychaudhuri,
  Phys.\ Rev.\ D {\bf 86}, 051302 (2012)
  [arXiv:1204.5619 [hep-ph]];
  B.~Adhikary, A.~Ghosal and P.~Roy,
  Int.\ J.\ Mod.\ Phys.\ A {\bf 28},  1350118 (2013)
  arXiv:1210.5328 [hep-ph];
  D.~Aristizabal Sierra, I.~de Medeiros Varzielas and E.~Houet,
  Phys.\ Rev.\ D {\bf 87}, 093009 (2013)
  [arXiv:1302.6499 [hep-ph]];
  R.~Dutta, U.~Ch, A.~K.~Giri and N.~Sahu,
  Int.\ J.\ Mod.\ Phys.\ A {\bf 29}, 1450113 (2014)
  arXiv:1303.3357 [hep-ph];
  L.~J.~Hall and G.~G.~Ross,
  JHEP {\bf 1311}, 091 (2013)
arXiv:1303.6962 [hep-ph];
  T.~Araki,
  PTEP {\bf 2013},  103B02 (2013)
arXiv:1305.0248 [hep-ph];
  M.~-C.~Chen, J.~Huang, K.~T.~Mahanthappa and A.~M.~Wijangco,
  JHEP {\bf 1310}, 112 (2013)
  [arXiv:1307.7711] [hep-ph].
  S.~Pramanick and A.~Raychaudhuri,
  Phys.\ Rev.\ D {\bf 88},  093009 (2013)
  [arXiv:1308.1445 [hep-ph]];
  B.~Brahmachari and P.~Roy,
  JHEP {\bf 1502}, 135 (2015)
  [arXiv:1407.5293 [hep-ph]].





\bibitem{katrin} 
  M.~Haag [KATRIN Collaboration],
  PoS EPS {\bf -HEP2013}, 518 (2013).

\bibitem{0nubeta} 
  W.~Rodejohann,
  Int.\ J.\ Mod.\ Phys.\ E {\bf 20}, 1833 (2011)
  [arXiv:1106.1334 [hep-ph]].











\end{thebibliography}
\end{document}